\documentclass{aa}  

\usepackage{graphicx}
\usepackage{txfonts}
\usepackage{natbib} 
\bibpunct{(}{)}{;}{a}{}{,} 

\usepackage{verbatim} 
\usepackage{hyperref}   
\hypersetup{colorlinks=true,linkcolor=blue,citecolor=blue,filecolor=blue,urlcolor=blue}

\usepackage{makeidx}
\makeindex

\begin{document}

   \title{Testing the asteroseismic estimates of stellar radii with surface brightness-colour relations and {\it Gaia} DR3 parallaxes}  

   \subtitle{II. Red giants and red clump stars from the {\it Kepler} catalogue}
\author{G. Valle \inst{1, 2}\orcid{0000-0003-3010-5252}, M. Dell'Omodarme \inst{1}\orcid{0000-0001-6317-7872}, P.G. Prada Moroni
        \inst{1,2}\orcid{0000-0001-9712-9916}, S. Degl'Innocenti \inst{1,2}\orcid{0000-0001-9666-6066}
}
\titlerunning{SBCR radii for Kepler sample}
\authorrunning{Valle, G. et al.}

\institute{
        Dipartimento di Fisica "Enrico Fermi'',
        Universit\`a di Pisa, Largo Pontecorvo 3, I-56127, Pisa, Italy\\
        \email{valle@df.unipi.it}
        \and
        INFN,
        Sezione di Pisa, Largo Pontecorvo 3, I-56127, Pisa, Italy
}

   \date{Received ; accepted }

  \abstract
{A recent investigation highlighted peculiar trends between the radii derived from surface brightness–colour relations (SBCRs) combined with {\it Gaia} DR3 parallaxes with respect to asteroseismic scaling relation radii from K2 data. 

}
{ {\it Kepler} data differ from K2 data in many aspects. We investigated on the robustness of the results based on Kepler data.
}
{
We cross-matched asteroseismic and astrometric data for over 12,000 red giant branch and red clump stars from
the end-of-mission {\it Kepler} catalogue with the {\it Gaia} DR3 and TESS Input Catalogue v8.2 to obtain precise parallaxes, $V$- and $K$-band magnitudes, and $E(B - V)$ colour excesses.
Two well-tested SBCRs from the literature were adopted to estimate stellar radii. 
}
{
The analysis confirmed that SBCR and asteroseismic radii agree very well. The overall differences are only 1-2\% depending on the adopted SBCR. The dispersion of 7\% was about two-thirds of what was found for K2-based data. 
As a difference from the K2-based investigation, the ratio of SBCRs-to-asteroseismic radii did not depend on the metallicity [Fe/H].
Moreover, the intriguing decreasing trend with [$\alpha$/Fe] of the radius ratio for massive stars that was  observed in K2 data was absent ind {\it Kepler} data.
The SBCR radii are systematically higher than asteroseismic estimates by 5\% for stars with masses below 1.0 $M_{\sun}$.
}
{
The SBCRs have proven to be a highly effective tool for estimating radii with a precision comparable to that obtained from asteroseismology, but at a significantly lower observational cost. Moreover, the superior concordance of {\it Kepler}-derived radii with SBCR measurements and the absence of the discrepancies observed in the K2-derived radii suggest the existence of underlying systematic errors that impact specific mass and metallicity regimes within the K2 dataset.
} 

   \maketitle

\section{Introduction}\label{sec:intro}

This paper is a follow-up of \citet{Valle2024raggi}\defcitealias{Valle2024raggi}{Paper I} (hereafter \citetalias{Valle2024raggi}), who investigated the agreement between stellar radii derived from surface brightness–colour relations (SBCRs), combined with accurate parallaxes from {\it Gaia} DR3, with those from asteroseismic scaling relations for evolved stars.  
We aim to replicate the analysis presented in this paper based on K2 data, but using the catalogue by \citet{Yu2018}, who presented asteroseismic observations of over 16,000 red giant branch (RGB) and red clump (RC) stars from the end-of-mission long-cadence data for the {\it Kepler} satellite \citep{Borucki2010}. The \citet{Yu2018} dataset also classifies the stellar evolutionary phases (RGB, RC, and unknown) and the scaling relation masses and radii for all the stars. The reported values from the scaling relations were corrected according to the \citet{Sharma2016} prescriptions, which account for the stellar evolutionary phase. The same corrections were included in the K2 data analysed in \citetalias{Valle2024raggi}.

The K2 and {\it Kepler} datasets exhibit significant differences in several aspects, including the asteroseismic pipelines employed for the analysis, the correction of systematics inherent to K2 data \citep{Zinn2022, Lund2024}, the target selection criteria, the line of sight, and the shorter duration of K2 light curves compared to {\it Kepler} \citep[see e.g.][]{Zinn2022, Stasik2024}. 
By comparing the results from the \citet{Yu2018} catalogue with those in \citetalias{Valle2024raggi}, which were derived using the APO-K2 catalogue dataset \citep{Stasik2024}, we aim to investigate some of the trends reported in \citetalias{Valle2024raggi} and gain valuable insights.

\section{Adopted SBCRs and data selection}

The surface brightness,  $S_{\lambda}$, of a star is linked to its limb-darkened angular diameter, $\theta,$ and its apparent magnitude
corrected for the extinction, $m_{\lambda0}$. In the $V$ band, $S_V$ is defined as
\begin{equation}
 S_V = V_0 + 5 \log \theta,   \label{eq:sv1}
\end{equation}
where $V_0$ is the $V$-band magnitude corrected for extinction. It follows from Eq.~(\ref{eq:sv1}) that
\begin{eqnarray}
\theta &=& 10^{0.2 \; (S_V - V_0)} \label{eq:theta} \\
    r &=& 0.5 \; d \; \theta, \label{eq:r}
\end{eqnarray}
where $d$ is the heliocentric distance of the star, and $r$ is an estimate of the stellar linear radius.

As in \citetalias{Valle2024raggi}, we adopted two different SBCRs. The first, proposed by \citet{Pietrzynski2019}, {which was calibrated with the \citet{Gallenne2018} giant star sample,} was
\begin{equation}
S_V^a = 1.330  [(V - K)_0 - 2.405] + 5.869 \; {\rm mag},
\end{equation}
where $(V-K)_0$ is the colour corrected for the reddening. This relation was fitted in the range $ 2.0<(V - K)_0<2.8 $~mag.
The second adopted SBCR was proposed by \citet{Salsi2021},
\begin{equation}
S_V^b = 1.22 (V - K)_0 + 2.864 \; {\rm mag}.
\end{equation}
This relation (from Table 5 in \citealt{Salsi2021} for stars of the spectral class F5/K7-II/III) is valid in the range $ 1.8<(V - K)_0<3.9$~mag.

The \citet{Yu2018} catalogue does not provide $V$ and $K$ magnitudes or {astrometric information, such as the parallaxes}. 
To obtain $V$- and $K_s$-band magnitudes and $E(B-V)$, we cross-matched the {\it Kepler} catalogue with the TESS Input Catalogue (TIC) v8.2. The TIC adopts the three-dimensional empirical dust maps from the Panoramic Survey Telescope and Rapid Response System \citep{Green2018}, with a recalibration coefficient of 0.884 applied to obtain  $E(B-V)$ values, as prescribed by \citet{Schlafly2011}.
The parallaxes were obtained from a cross-match to the {\it Gaia} DR3 dataset \citep{Gaia2021}, and were corrected according to the {\it Gaia} zero-point \citep{Lindegren2021}.

{The \citet{Yu2018} catalogue provides asteroseismic radius estimates for all stars, derived from corrected scaling relations. These corrections were computed using data from \citet{Sharma2016}, which incorporate adjustments based on the evolutionary phase (RGB or RC) of each star \citep[see][for details]{Yu2018}.}

The data in the catalogue were subjected to a selection procedure to reject apparent outliers and to restrict the data to a metallicity range in which asteroseismic scaling relations are most reliable. 
{All the stars are in the giant evolutionary phase, which spans a range of $\log g$ for about 3.3 dex to 1.6 dex, and an effective temperature lower than 5500 K.}
Stars with [Fe/H] > $-1.0$ dex were retained in the sample. Stars in the RGB with a mass lower than 0.75 $M_{\sun}$ were rejected as artefacts because single stars with a mass this low cannot be on the RGB given their long evolutionary timescale. Stars outside the range of colour for the \citet{Pietrzynski2019} SBCR were excluded to ensure a common validity range of the two SBCRs. Finally, stars with a relative error in the parallax greater than 0.1 were rejected. This allowed us to rely on distances obtained using the inverse parallax \citep{Bailer2021, Fouesneau2023}.  
The final sample comprised 12,492 stars, 6,034 of which were on the RGB and 6,458 of which were in the RC phase. This samples is twice as large as that in \citetalias{Valle2024raggi}, and the RC sample is about three times larger.

\section{SBCRs to asteroseismic radii comparisons}\label{sec:results}

\begin{figure*}
        \centering
        \includegraphics[width=17cm]{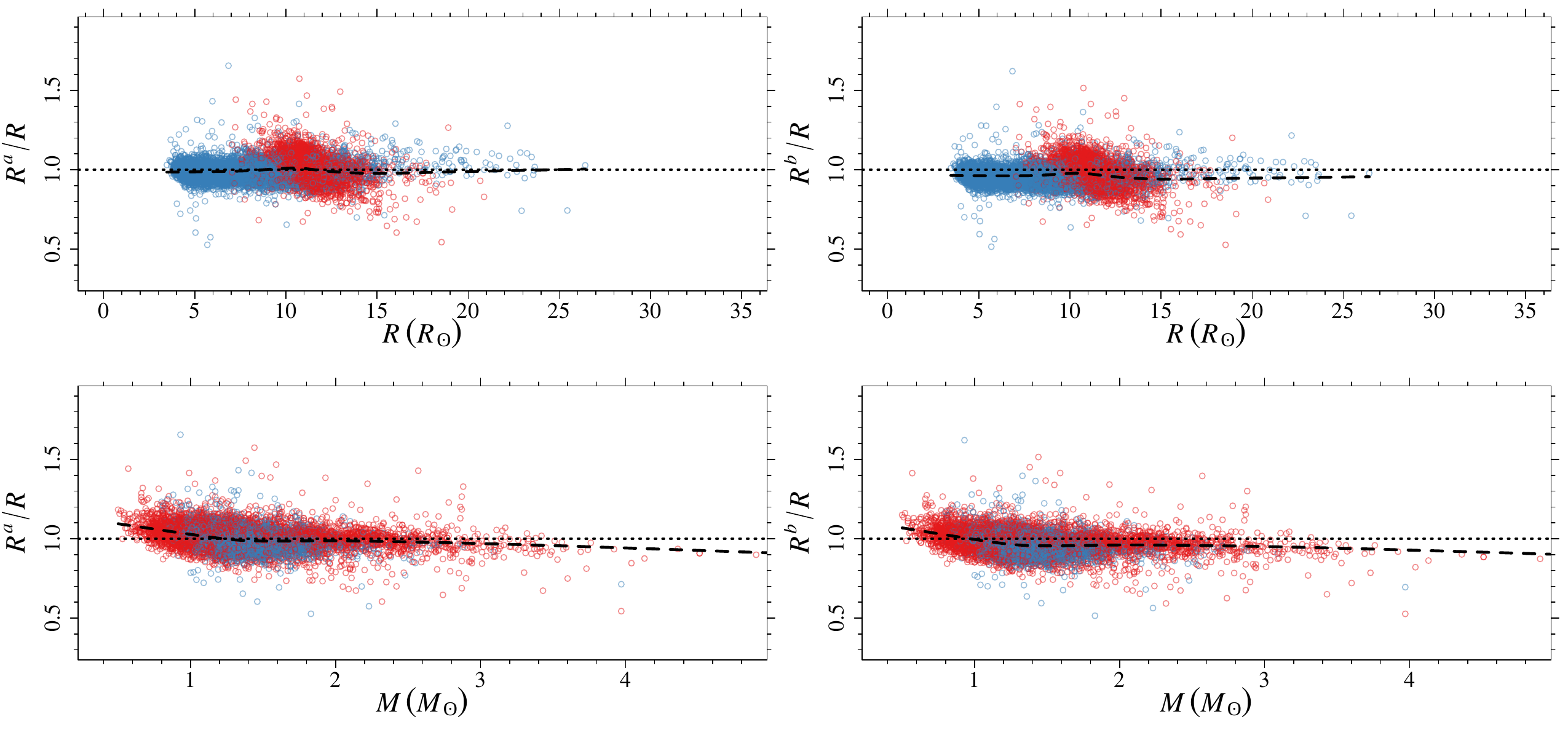}
        \caption{Ratio of the radii from SBCRs and from asteroseismology as a function of masses and radii estimated through the scaling relations.   {\it Top row, left}: Radius ratio using the \citet{Pietrzynski2019} SBCR. The red and blue points correspond to RC and RGB stars, respectively. The dashed black line is a LOWESS smoother of data, and the dotted line serves as a visual aid. 
        {\it Top row, right}: Same as in the left panel, but using the \citet{Salsi2021} SBCR.  
        {\it Bottom row, left}: Same as in the left panel of the top row, but as a function of the asteroseismic mass.
        {\it Bottom row, right}: Same as in the left panel, but using the \citet{Salsi2021} SBCR.  
        }
        \label{fig:RM}
\end{figure*}

For all stars in the final sample, the linear radii were calculated using Eq.~(\ref{eq:r}). The angular diameter, $\theta$, was determined using the SBCR calibration from \citet{Pietrzynski2019} and \citet{Salsi2021}, resulting in estimates denoted $R^a$ and $R^b$, respectively.
Figure~\ref{fig:RM} shows the scatter plot of the ratios $R^a/R$ and $R^b/R$, where $R$ is the radius from the scaling relations as a function of the scaling relation radii and masses. The results largely confirm the findings of \citetalias{Valle2024raggi} and show that SBCR and scaling relation radii for the {\it Kepler} sample agree very well. The median differences across the entire sample were 1.0\% and 2.2\% for the \citet{Pietrzynski2019} and \citet{Salsi2021} SBCRs, respectively. These values are similar to the 1.2\% and $-2.5\%$ found in the K2-based analysis.
However, in contrast to what we found in \citetalias{Valle2024raggi} for the K2 dataset, we did not observe a systematic negative distortion in $R^a/R$ and $R^b/R$ for stars with $R > 15$ $R_{\sun}$ in the {\it Kepler} dataset. Additionally, with standard deviations of approximately 7\% for both SBCRs, the dispersion of the radii ratios was lower than that reported for K2 by \citetalias{Valle2024raggi}, which was about 10\%. 
For comparison, this dispersion is roughly twice the magnitude of the relative errors on the radius estimates of the individual techniques. Specifically, the relative errors for  the SBCR radius determinations are 3.2\% and 4.0\% for the {\it Kepler} and K2 datasets, respectively. These values are very close to the relative errors in asteroseismic radii, which are 3.7\% and 4.3\% for {\it Kepler} and K2 datasets.       

\begin{figure}
        \centering
        \resizebox{\hsize}{!}{\includegraphics{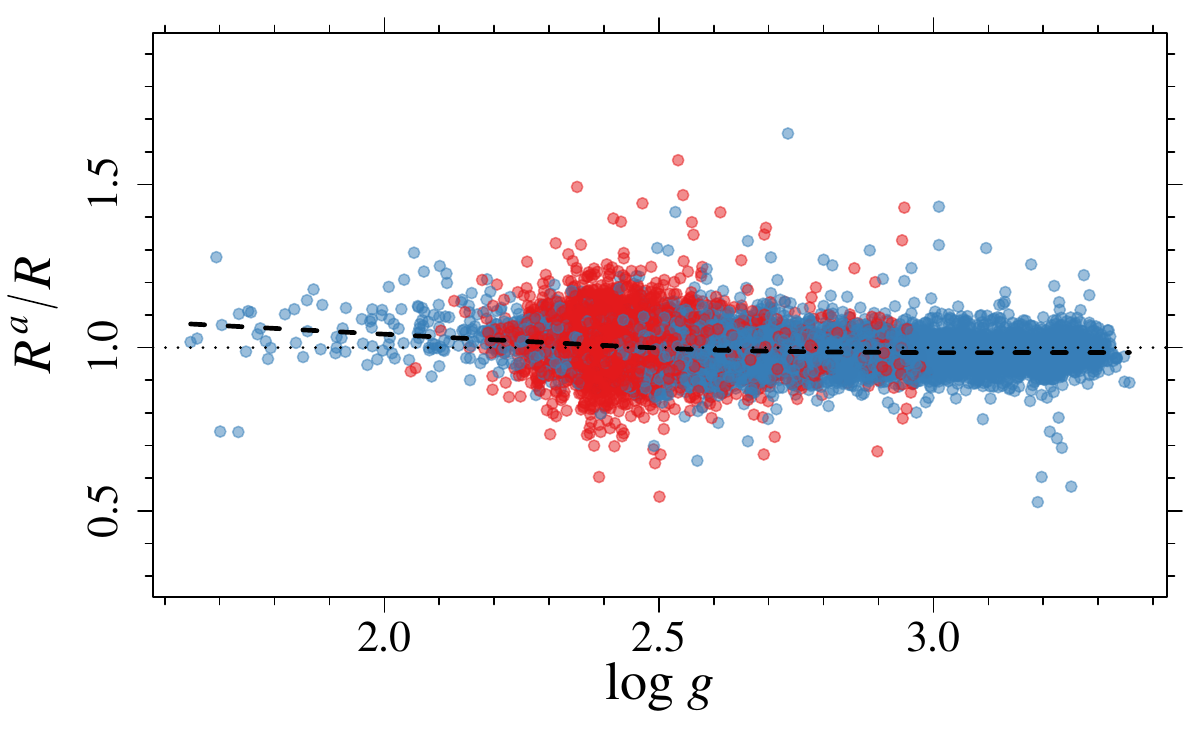}}
        \caption{Ratio of the radii from \citet{Pietrzynski2019} SBCR and from asteroseismology as a function of $\log g$. The colour code is the same as in Fig.~\ref{fig:RM}. The dashed black line is a LOWESS smoother of RGB data, and the dotted one serves as a visual aid.  }
        \label{fig:logg}
\end{figure}

\begin{figure}
        \centering
        \resizebox{\hsize}{!}{\includegraphics{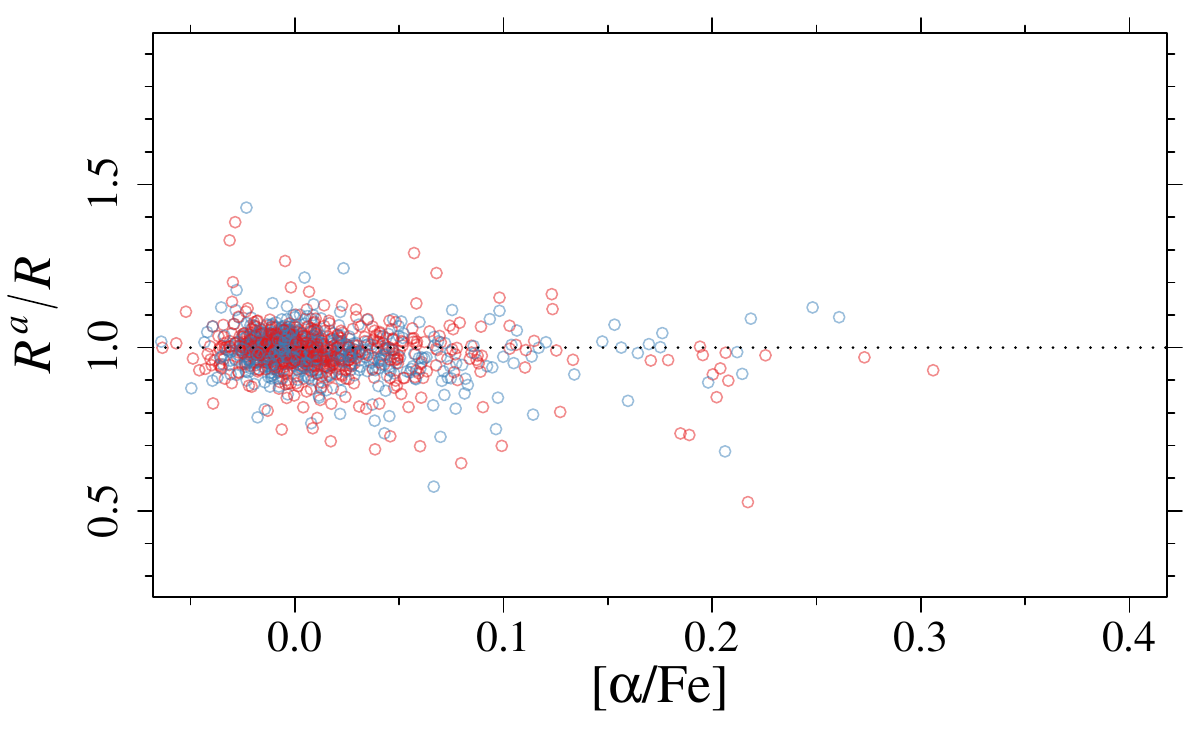}}
        \caption{Ratio of the radii from \citet{Pietrzynski2019} SBCR and from asteroseismology as a function of [$\alpha$/Fe]. Only stars with $M > 1.7$ $M_{\sun}$ are shown. The colour code is the same as in Fig.~\ref{fig:RM}. }
        \label{fig:alpha}
\end{figure}

\begin{figure}
        \centering
        \resizebox{\hsize}{!}{\includegraphics{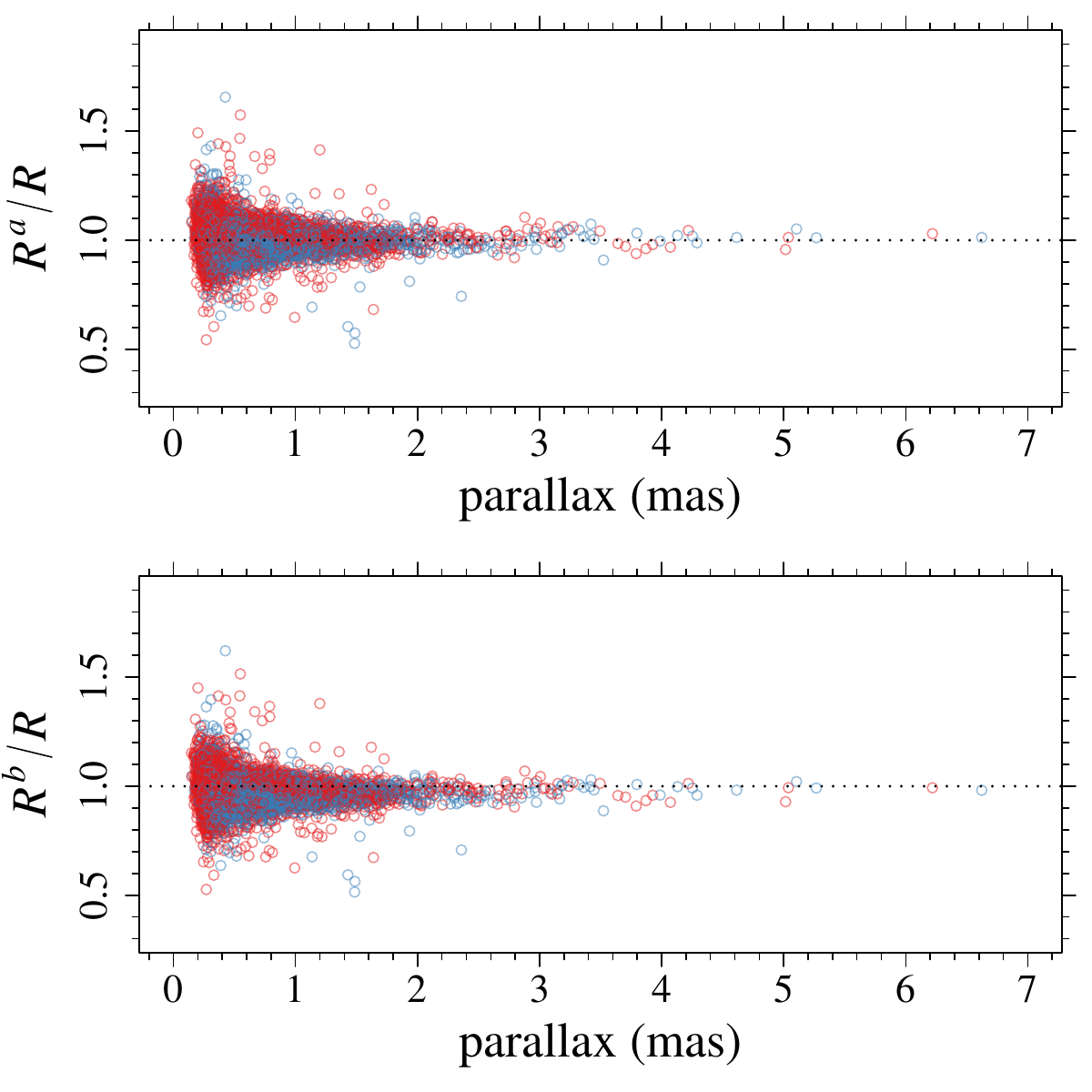}}
        \caption{Ratio of the radii from SBCRs and from asteroseismology as a function of the parallax. {\it Top}: radii ratio for the  \citet{Pietrzynski2019} SBCR. {\it Bottom}: Same as in the top panel, but for  the \citet{Salsi2021} SBCR. The colour code is the same as in Fig.~\ref{fig:RM}. }
        \label{fig:R-para}
\end{figure}

The relation between SBCR-based radii and asteroseismic mass $M$ (bottom row in Fig.~\ref{fig:RM}) exhibits both similarities and differences compared to the results of \citetalias{Valle2024raggi}. Similarly to what was observed in that paper, the SBCR method consistently yields larger radii at lower masses. The comparison of the median radii ratios for the two SBCRs reveals a difference of approximately 5\% between the $M < 1.0$ $M_{\sun}$ and $M > 1.0$ $M_{\sun}$ subsamples.
The discrepancy at the low-mass end is partly attributable to the presence of 170 RC stars with masses below 0.75 $M_{\sun}$  whose SBCR-based radii exceed the scaling relation radii by approximately 5\%. As noted in \citetalias{Valle2024raggi}, these stars may be the result of mass loss and cannot be dismissed as artefacts.
\citetalias{Valle2024raggi} also proposed that the development of a substantial helium core during the RGB evolution  might contribute to this trend. {\it Kepler} data support this hypothesis because a tendency for radii ratios greater than one is observed at lower surface gravity $\log g < 2.5$ (Fig.~\ref{fig:logg}). Specifically, $R^a$ was higher by 3.2\% on average than $R$ in the late-RGB evolutionary phase. However, this difference vanished in the lower RGB part. Notably, this behavior did not correlate with stellar mass, unlike the findings of \citetalias{Valle2024raggi}. The trends of $R^a/R$ and $R^b/R$ with $\log g$ were independent of the mass, while \citetalias{Valle2024raggi} reported that the trend was restricted to stars with $M \lesssim 0.95$ $M_{\sun}$.
However, the contribution of late-RGB and low-mass RC stars alone is insufficient to explain the observed discrepancy in the radius ratios. 
As discussed in \citetalias{Valle2024raggi}, this discrepancy might originate from the paucity of stellar targets below 0.9 $M_{\sun}$ adopted for SBCR calibrations, as well as from an inherent bias in the asteroseismic relations.
The theoretical analysis by \citet{Salsi2022} reported negligible differences between $\log g = 3$ and $\log g = 0$ in the predicted SBCR radii.
Further research into this area could provide valuable insights.

For stars with masses greater than 1.5 $M_{\sun}$, our results differ from those of \citetalias{Valle2024raggi}. Specifically, we do not observe the significant number of stars with radius ratios as low as 0.5 they reported. Consequently, the decreasing trend in $R^a/R$ and $R^b/R$ with $M$ is not confirmed with the data from \citet{Yu2018}.

\citetalias{Valle2024raggi} observed a peculiar dependence of the radius ratios on [$\alpha$/Fe] that was restricted to massive stars ($M \gtrsim 1.7$ $M_{\sun}$). They reported a significant negative linear trend of approximately $-1.0$ and $-1.5$ per dex for RGB and RC stars, respectively. To investigate this trend in our sample, we obtained the [$\alpha$/Fe] by cross-matching our dataset with the APOGEE DR17 catalogue \citep{Abdurrouf2022}. We identified 7,562 matching stars (3,368 RGB and 4,194 RC). This data source is the same as was adopted in the APO-K2 catalogue investigated in the previous work. 
Interestingly, we did not reproduce the [$\alpha$/Fe] trend based on the asteroseismic estimates from the {\it Kepler} mission. As shown in Fig.~\ref{fig:alpha}, even when we restricted our analysis to the 1,161 stars with $M > 1.7$ $M_{\sun}$, the radius ratio remains consistently uniform across different [$\alpha$/Fe] values.

\begin{figure}
        \centering
        \resizebox{\hsize}{!}{\includegraphics{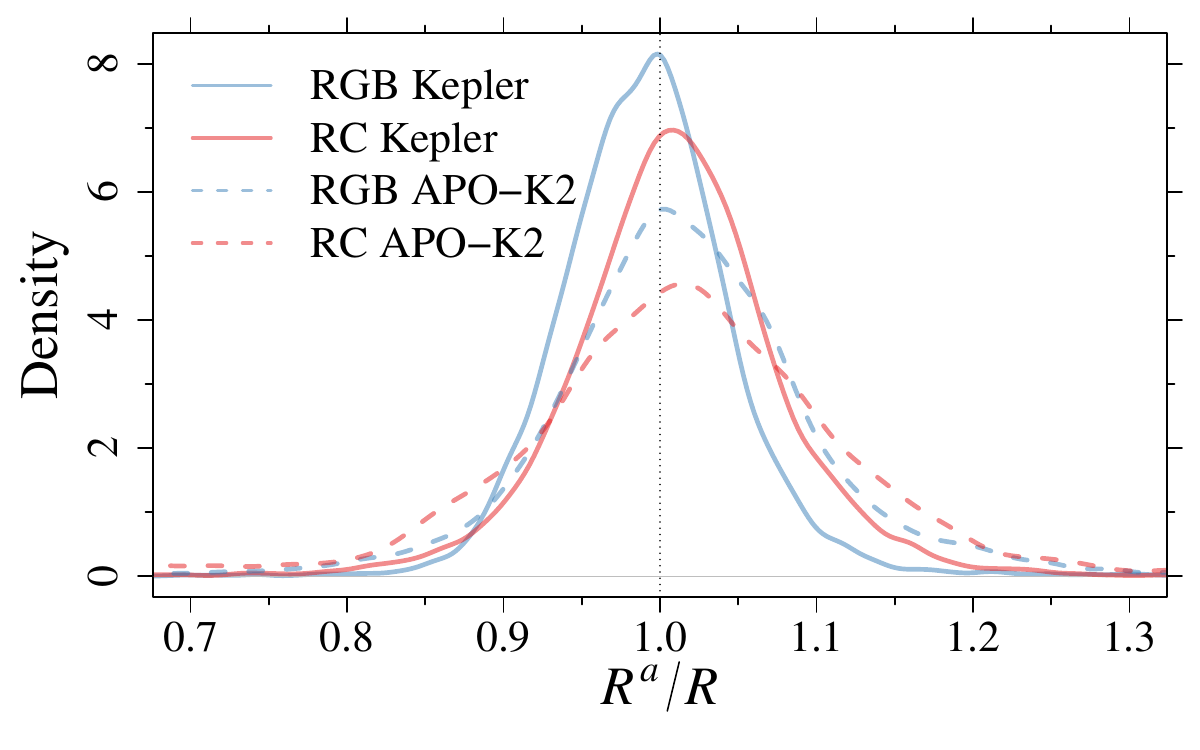}}
        \caption{Kernel density estimators for the ratio of radii derived from SBCR and asteroseismology, using the \citet{Pietrzynski2019} SBCR. The solid blue and red lines represent the results for RGB and RC stars, respectively, based on the \citet{Yu2018} dataset. The dashed lines correspond to the results obtained using the APO-K2 dataset. }
        \label{fig:cfr}
\end{figure}

Unlike \citetalias{Valle2024raggi}, we found no correlation between radius ratios and metallicity [Fe/H]. Robust regression models indicate a slope of 0.0\% and $-1.0\%$ for $R^a/R$ and $R^b/R$, respectively. These results align well with theoretical investigations by \citet{Salsi2022}, who suggested a minimum impact of the metallicity at an effective temperature of approximately 5000 K.

Similar to \citetalias{Valle2024raggi}, the agreement between SBCR-based radii and asteroseismic radii is significantly better for nearby stars with accurate parallax measurements (Fig.~\ref{fig:R-para}). For parallax values exceeding 2.5 mas, $R^a$ was higher by 1.2\% on average than $R$, while $R^b$ was 1.3\% lower than $R$.

Figure~\ref{fig:cfr} compares the kernel density estimators \citep{simar} of $R^a/R$ for RGB and RC stars from the {\it Kepler} and APO-K2 datasets. As noted earlier, the results based on {\it Kepler} data exhibit a lower dispersion, with the RGB density function centred around the unbiased value of one. The APO-K2 RGB density estimator shows a distinct positive skewness that is not observed in the other densities. Both RC datasets present a slight positive bias in the mode, with peaks around 1.02. The agreement between the two datasets is very satisfactory overall and supports the use of SBCR radii, which also provide the additional advantage of achieving comparable results with shorter observing times.

\section{Conclusions}\label{sec:conclusions}

We compared the radii derived from asteroseismic scaling relations with those obtained from SBCRs combined with {\it Gaia} DR3 parallaxes. For this investigation, we used the {\it Kepler} data from the \citet{Yu2018} catalogue, thereby expanding upon the analysis conducted in \citetalias{Valle2024raggi}, where we used the APO-K2 catalogue \citet{Stasik2024}.
The differences between these catalogues allowed us to assess the robustness of the trends identified in \citetalias{Valle2024raggi} using a dataset of approximately 12,000 stars, which is twice the size of the sample analysed in the previous work.

Following the method of \citetalias{Valle2024raggi}, we employed SBCRs  from \citet{Pietrzynski2019} and \citet{Salsi2021}. Information regarding colour excess and magnitudes in the $V$ and $K_s$ bands was obtained by cross-matching the \citet{Yu2018} catalogue with {\it Gaia} DR3 and TIC v8.2.

The excellent agreement between SBCR radii and asteroseismic estimates was confirmed for both SBCRs we analysed. This agreement is quite relevant because asteroseismic scaling relations and SBCRs rely on fundamentally different information.
The median radius differences were 1.0\% and 2.2\% for the \citet{Pietrzynski2019} and the \citet{Salsi2021} SBCRs, respectively. In contrast to \citetalias{Valle2024raggi}, we did not observe any bias in the ratio of SBCR-based radii to asteroseismic radii, even for higher radius values.
The dispersion of the radius ratios was approximately 7\% for both SBCRs, which is about 70\%  of the value reported by \citetalias{Valle2024raggi}.
Moreover, we found no dependence of radius ratios on [Fe/H], while \citetalias{Valle2024raggi} reported an unexpected increasing trend of about 4\% to 6\% per dex.

The most notable differences compared to \citetalias{Valle2024raggi} arise from the relation between radius ratios and asteroseismic mass. \citetalias{Valle2024raggi} identified a dichotomous behaviour in the radius ratios for high-mass stars: while some of them had a ratio close to one, for others, the ratios were as low as 0.5. However, when {\it Kepler} data were adopted for the analysis, the radius ratios are consistent with one for the whole subsample of massive stars. Conversely, we do confirm the discrepancy between SBCR and asteroseismic radii for stars with $M < 1.0$ $M_{\sun}$, with SBCR estimates exceeding asteroseismic ones by approximately 5\%.

\citetalias{Valle2024raggi}  reported an intriguing dependence between radius ratios and $\alpha$ enhancement in massive stars ($M \gtrsim 1.7$ $M_{\sun}$). To assess the robustness of this trend in the {\it Kepler} sample, we cross-matched the \citet{Yu2018} dataset with the APOGEE DR17 catalogue. Interestingly, we found that this trend also disappears when {\it Kepler} data wre used for the comparison.

The better agreement between SBCR radii and {\it Kepler}-based results, with respect to K2-based results is expected considering the shorter duration of K2 light curves compared to Kepler \citep{Stasik2024} light curves and the need for systematic corrections in K2 data \citep{Lund2024}. Since SBCR radii were estimated consistently with \citetalias{Valle2024raggi}, the differences from previous results stem from inherent differences in the asteroseismic data from \citet{Stasik2024} and \citet{Yu2018}. The processing of asteroseismic data and the origin of the effective temperatures are different between \citet{Yu2018} and the APO-K2 catalogue \citet{Stasik2024}.
The disappearance of the suspicious and unexpected trends seen in \citetalias{Valle2024raggi} seems to suggest that they are artefacts that are caused by biases in certain mass and metallicity regimes in the K2 dataset.
Further research in this area is encouraged to understand whether a systematic exists between the {\it Kepler} and K2-based datasets.
Beyond the absence of the suspicious trends reported in \citetalias{Valle2024raggi}, the most interesting finding of this study is that we confirm the reliability of radii estimated from SBCRs. Given the significantly lower requirements in terms of observational time, instrumentation, and financial resources, SBCR radii appear to be a viable alternative to asteroseismology-based radii.

{While this paper and \citetalias{Valle2024raggi} focused on the agreement of SBCRs with asteroseismic radii in giant stars, the applicability of SBCRs extends beyond this specific evolutionary phase. Numerous studies have calibrated SBCRs for main-sequence stars in various mass ranges. While some calibrations for dwarf stars used classic photometric bands such as $V$, $I$, $J$, $H$, and $K$ \citep[e.g.][]{Kervella2004, Boyajian2012, Adams2018}, others leveraged the higher precision of {\it Gaia} photometry \citep[e.g.][]{Salsi2021, Graczyk2021, Kiman2024}. A comparison of asteroseismology and SBCR-derived radii for dwarf stars would be valuable to investigate potential systematic differences attributable to the stellar evolutionary stage.}

\begin{acknowledgements}
G.V., P.G.P.M. and S.D. acknowledge INFN (Iniziativa specifica TAsP) and support from PRIN MIUR2022 Progetto "CHRONOS" (PI: S. Cassisi) finanziato dall'Unione Europea - Next Generation EU.
\end{acknowledgements}

\bibliographystyle{aa}
\bibliography{biblio}

\end{document}